\documentclass[12pt]{article}

\usepackage[paper=a4paper,left=24mm,right=24mm,top=24mm,bottom=34mm]{geometry}
\usepackage{amsfonts,amsmath,amssymb}
\usepackage{bbm,booktabs,graphicx,multirow,setspace}
\usepackage{color}
\usepackage[hyphens]{url}
\usepackage[english]{babel}
\usepackage[utf8]{inputenc}
\usepackage[round]{natbib}
\usepackage{authblk}

\newcommand{\myI}{\ensuremath{\mathbbm{1}}}

\newcommand{\HRES}{{\rm HRES}}
\newcommand{\CNT}{{\rm CNT}}
\newcommand{\ENS}{{\rm ENS}}

\newcommand{\ww}{$^{--}$}
\newcommand{\wc}{$^{-}$}
\newcommand{\bb}{$^{++}$}
\newcommand{\bc}{$^{+}$}

\title{Skill of global raw and postprocessed ensemble predictions of rainfall over northern tropical Africa}

\author[1,2]{Peter Vogel \thanks{peter.vogel3@kit.edu}}
\author[1]{Peter Knippertz}
\author[1]{Andreas H.~Fink}
\author[1]{Andreas Schlueter}
\author[2,3]{Tilmann Gneiting}
\affil[1]{Institute of Meteorology and Climate Research, Karlsruhe Institute of Technology}
\affil[2]{Institute for Stochastics, Karlsruhe Institute of Technology}
\affil[3]{Heidelberg Institute for Theoretical Studies}

\begin{document}

\maketitle


\begin{abstract}
 Accumulated precipitation forecasts are of high
 socioeconomic importance for agriculturally dominated societies in
  northern tropical Africa.  In this study, we analyze the performance
  of nine operational global ensemble prediction systems (EPSs)
  relative to climatology-based forecasts for 1 to 5-day accumulated
  precipitation based on the monsoon seasons 2007--2014 for three
  regions within northern tropical Africa.  To assess the full
  potential of raw ensemble forecasts across spatial scales, we apply
  state-of-the-art statistical postprocessing methods in form of
  Bayesian Model Averaging (BMA) and Ensemble Model Output Statistics
  (EMOS), and verify against station and spatially aggregated, 
  satellite-based gridded observations.  Raw ensemble forecasts are
  uncalibrated, unreliable, and underperform relative to climatology,
  independently of region, accumulation time, monsoon season, and
  ensemble.  Differences between raw ensemble and climatological
  forecasts are large, and partly stem from poor prediction for low
  precipitation amounts.  BMA and EMOS postprocessed forecasts are
  calibrated, reliable, and strongly improve on the raw ensembles, but
  -- somewhat disappointingly -- typically do not outperform
  climatology.  Most EPSs exhibit slight improvements over the period
  2007--2014, but overall have little added value compared to
  climatology.  We suspect that the parametrization of convection is a
  potential cause for the sobering lack of ensemble forecast skill in
  a region dominated by mesoscale convective systems.
\end{abstract}

%
%
\section{Introduction}  \label{sec:Introduction}
The bulk of precipitation in the Tropics is related to moist
convection, in contrast to the frontal-dominated extratropics.  Due to
the small-scale processes involved in the triggering and growth of
convective systems, quantitative precipitation forecasts are known to
have overall poorer skills in tropical latitudes \citep{Haiden2012}.
This can be monitored in quasi-real time on the World Meteorological
Organisation (WMO) Lead Centre on Verification of Ensemble Prediction
System website (\url{http://epsv.kishou.go.jp/EPSv}) by comparing
deterministic and probabilistic skill scores for 24-hour precipitation
forecasts for the $20^\circ$N--$20^\circ$S tropical belt with those
for the northern and southern hemisphere extratropics.  There are
hints that precipitation and cloudiness forecasts in the Tropics show
enhanced skill during regimes of stronger synoptic-scale forcing
\citep{Sohne2008, Davis2013, vanderLinden2017} or in regions of 
orographic forcing \citep{Lafore2017}, but large parts of the tropical
land masses are dominated by convection that initiates from
small-scale surface and boundary layer processes and sometimes
organizes into mesoscale convective systems (MCSs).  The latter
depends mostly on the thermodynamic profile and vertical wind shear.

In this context, northern tropical Africa, particularly the semi-arid
Sahel, can be considered a region where precipitation forecasting is
particularly challenging.  The area consists of vast flatlands, MCSs
during boreal summer provide the bulk of the annual rainfall
\citep{Mathon2002, Fink2006, Houze2015}, and convergence lines in the
boundary layer or soil moisture gradients at the km-scale can act as
triggers for MCSs \citep{Lafore2017}.  Sahelian MCSs often take the
form of meridionally elongated squall lines with sharp leading edges
characterized by heavy rainfall.  Synoptic-scale African easterly
waves are known to be linked to squall line occurrence in the western
Sahel \citep{Fink2003} and lead to an enhanced skill of cloudiness
forecasts over West Africa \citep{Sohne2008}.

However, numerical weather prediction (NWP) models are known to have
an overall poor ability to predict rainfall systems over northern
Africa.  For example, the gain in skill by improved initial conditions
due to an enhanced upper-air observational network during the 2006
AMMA campaign \citep{Parker2008} was lost in NWP models after 24 hours
of forecast time, potentially due to the models' inability to predict
the genesis and evolution of convective systems \citep{Fink2011a}.

Given the substantial challenges involved in forecasting rainfall in
northern Africa, one might hope that ensemble prediction systems
(EPSs) provide a useful assessment of uncertainties and a more useful
forecast overall.  An ensemble is a set of deterministic forecasts,
created by changes in the initial conditions and/or the numerical
representation of the atmosphere \citep{Palmer2002}.  With clear
advantages of ensembles over single deterministic forecasts, EPSs are
now run at all major NWP centers, which led to the creation of the
TIGGE multi-model ensemble database \citep{Bougeault2010,
Swinbank2016}.  TIGGE contains forecasts from up to ten global EPSs,
with the ensemble of the European Centre for Medium-Range Weather
Forecasts (ECMWF) being the most prominent and most important
contributor \citep{Hagedorn2012}.  To our knowledge, this present
study is the first to rigorously and systematically assess the quality
of ensemble forecasts for precipitation over northern tropical Africa.
This is partly related to the fact that for this region ground
verification data from rain gauge observations are infrequent on the
Global Telecommunication System (GTS), the standard verification data
source for NWP centers.

Despite many advances in the generation of EPSs, ensembles share
structural deficiencies such as dispersion errors and biases.
Statistical postprocessing addresses these deficiencies and thereby
allows assessing the full potential of ensemble forecasts
\citep{Gneiting2005a}.  Additionally, it enables fair comparisons
between different spatial scales, such as model gridboxes and point
observations.  The correction of systematic forecast errors is based
on (distributional) regression techniques and, depending on the need
of the user, several approaches are at hand \citep{Schefzik2013, 
Gneiting2014a}.  \cite{Hamill2004} and \cite{Wilks2009} proposed and
extended logistic regression techniques, which yield probabilistic
forecasts for the exceedance of thresholds.  Here we will for the
first time explore whether established methods such as Bayesian Model
Averaging \citep[BMA,][]{Raftery2005a} and Ensemble Model Output
Statistics \citep[EMOS,][]{Gneiting2005}, which provide complete
probabilistic quantitative precipitation forecasts, can improve
precipitation forecasts for Africa.

The ultimate goal of this paper is to provide an exhaustive assessment
of our current ability to predict rainfall over northern tropical
Africa, considering the skill of raw and postprocessed forecasts from
TIGGE.  Any skill, if existing, would be expected to come from
resolved large-scale forcing processes as mentioned above.  We examine
accumulation periods of 1- to 5-days for the monsoon seasons
2007--2014 and verify against about 21,000 daily rainfall observations
from 132 rain gauge stations and satellite-based gridded precipitation
observations.  Section \ref{sec:Data} introduces the TIGGE ensemble,
and the station and satellite-based observations used for
verification.  Section \ref{sec:Methods} describes our benchmark
climatological forecast and methods for the evaluation of
probabilistic forecasts, and explains EMOS and BMA in detail.  Results
are presented in section \ref{sec:Results}, where we verify 1-day
accumulated ECMWF precipitation forecasts against station
observations.  This analysis is performed in particular depth and
serves as fundamental examplar.  We also evaluate ECWMF ensemble
forecasts at longer accumulation times and for spatial aggregations,
before turning to the analysis of all TIGGE sub-ensembles.
Implications of our findings and possible alternative methods for
forecasting precipitation over northern tropical Africa are discussed
in section \ref{sec:Discussion}.

\section{Data}  \label{sec:Data}

\subsection{Forecasts}

The TIGGE multi-model ensemble was set up as part of the THORPEX
programme in order to ``accelerate improvements in the accuracy of
1-day to 2-week high-impact weather forecasts for the benefit of
humanity'' \cite[p.~1060]{Bougeault2010}.  Since its start in October
2006, up to 10 global NWP centers have provided their operational
ensemble forecasts, which are accessible on a common $0.5^\circ \times
0.5^\circ$ grid.  \cite{Park2008} and \cite{Bougeault2010} discuss
objectives and the set-up of TIGGE, including the participating EPSs,
in great detail.  They also note early results using the TIGGE
ensemble, while \cite{Swinbank2016} report on achievements
accomplished over the last decade.  \cite{Hagedorn2012} find that a
multi-model ensemble composed of the four best participating TIGGE
EPSs, which include the ECMWF ensemble, outperforms
reforecast-calibrated ECMWF forecasts.  For the evaluation of NWP
precipitation forecast quality, TIGGE is the most complete and best
available data source for the period 2007--2014.  Table
\ref{tab:TIGGE} gives an overview of the nine participating TIGGE EPSs
that provide accumulated precipitation forecasts.

In addition to the separate evaluation of each participating TIGGE
sub-ensemble, we construct a reduced multi-model (RMM) ensemble.  For
each of the seven sub-ensembles available for the period 2008--2013,
the RMM ensemble uses the mean of the perturbed members, and the
control run, and in case of the ECMWF EPS furthermore the
high-resolution run, as individual contributors.  The RMM ensemble
therefore consists of 15 members and, as postprocessing performs an
implicit weighting of all contributions, a manual selection of
sub-ensembles as performed by \cite{Hagedorn2012} is not necessary.

Arguably, the ECMWF EPS is the leading one among the TIGGE
sub-ensembles \citep{Buizza2005, Hagedorn2012, Haiden2012}.  It
consists of a high-resolution (HRES) run, a control (CNT) run, and 50
perturbed ensemble (ENS) members.  The HRES and CNT runs are started
from unperturbed initial conditions and differ only in their
resolution.  The ENS members are started from perturbed initial
conditions and have the same resolution as the CNT run.
\cite{Molteni1996} and \cite{Leutbecher2008} describe generation and
properties of the ECMWF system in detail.

\begin{table}[t]
\caption{TIGGE sub-ensembles used in this study, with years of
  availability, number of ensemble members (number of perturbed
  members + control run + any high-resolution run), initialization
  time (UTC), and native grid(s) used in the period
  2007--2014.  \label{tab:TIGGE}}
\medskip
\centering
\resizebox{\textwidth}{!}{
\begin{tabular}{llllll}
\hline
Source & Acronym & Availability & Members & Init time &  Native grid(s) \\
\hline
China Meteorological Administration                   & CMA   & 2008--13    & 14+1    & 00 & TL213/T639 \\
Centro de Previs\~{a}o Tempo e Estudos Clim\'{a}ticos & CPTEC & 2008--14    & 14+1    & 00 & T126 \\
European Centre for Medium-Range Weather Forecasts    & ECMWF & 2007--14    & 50+1+1  & 00 & T399/T639 \\
Japan Meteorological Agency                           & JMA   & 2007--13/14 & 50/26+1 & 12 & TL159/TL319/TL479 \\
Korea Meteorological Administration                   & KMA   & 2011--14    & 16+1    & 00 & N320 \\
M\'{e}t\'{e}o France                                  & MF    & 2010--14    & 34+1    & 06 & TL798 \\
Meteorological Service of Canada                      & MSC   & 2008--14    & 20+1    & 00 & 0.45$^\circ$ uniform \\
National Centres for Environmental Prediction         & NCEP  & 2008--14    & 20+1    & 00 & T126 \\
UK Met Office                                         & UKMO  & 2007--13    & 23+1    & 00 & N144/N216/N400 \\
\hline
\end{tabular}
}
\end{table}

\subsection{Observations}

Despite multiple advances in satellite rainfall estimation, station
observations of accumulated precipitation remain a reliable and
necessary source of information.  However, the meteorological station
network in tropical Africa is sparse and clustered, and observations
of many stations are not distributed through the GTS.  The Karlsruhe
African Surface Station Database (KASS-D) contains precipitation
observations from a variety of networks and sources.  Manned stations
operated by African national weather services provide the bulk of the
24-hour precipitation data.  Due to long-standing collaborations with
these services and African researchers, KASS-D contains many
observations not available in standard, GTS-fed station databases.
Within KASS-D, 960 stations have daily accumulated (usually 06--06
UTC) precipitation observations.  

\begin{figure}[h]
\centering
\includegraphics[page = 1, width = \textwidth]{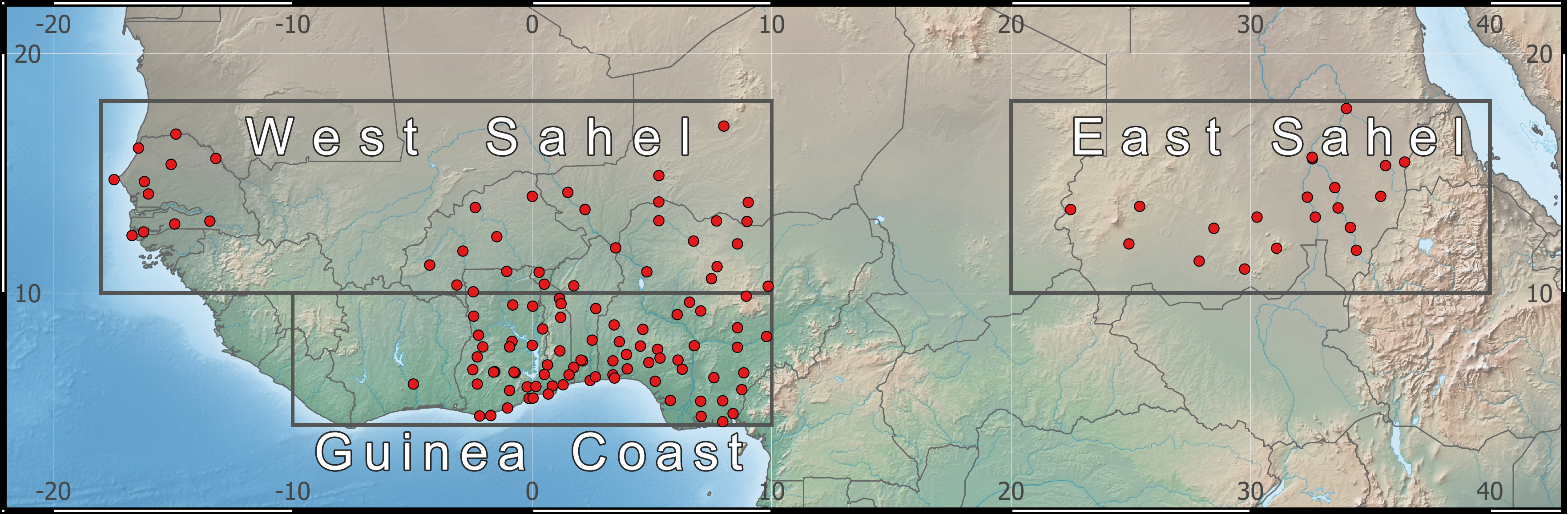}
\caption{Geographical overview of the study domain, with the locations
  of the observation stations (dots) within the three considered
  regions.  \label{fig:Africa}}
\end{figure}

After excluding stations outside the study domain, and removing sites
with less than 80\% available observations in any of the monsoon
seasons, the remaining 132 stations were subject to quality control,
as described in the Appendix, and passed these tests.  Based on their
rainfall climate \citep[e.g.][]{Fink2017} and geographic clustering,
the stations were assigned to three regions, as indicated in
Fig.~\ref{fig:Africa}: West Sahel, East Sahel, and Guinea Coast.

As NWP forecasts are issued for grid cells, the comparison of station
observations against gridded forecasts is fraud with problems.  To
allow for an additional assessment of forecast quality without a
gauge-to-gridbox comparison and for areas without station
observations, we use satellite-based, gridded precipitation estimates.
Based on recent studies, version 7 (and also version 6) of the
Tropical Rainfall Measuring Mission (TRMM) 3B42 gridded data set is
regarded the best available satellite precipitation product, despite a
small dry bias \citep{Roca2010, Maggioni2016, Engel2016}.

TRMM merges active measurements from the precipitation radar with
passive, radar-calibrated information from infrared as well as
microwave measurements \citep{Huffman2007}.  Based on monthly
accumulation sums, TRMM estimates are calibrated against nearby gauge
observations.  TRMM 3B42-V7 data are available on a $0.25^\circ \times
0.25^\circ$ grid with three hourly temporal resolution.

\subsection{Data preprocessing}

Based on 1-day accumulated station observations, we derive 2- to 5-day
accumulated precipitation observations by summing over consecutive
1-day observations.  As these cover the period from 06 UTC of the
previous day to 06 UTC of the considered day and as all TIGGE
sub-ensembles, except M\'{e}t\'{e}o France (MF), have initialization
times different from 06 UTC, we use the most recent run available at
that time, and adapt accordingly.  Specifically, for the sub-ensembles
initialized at 00 UTC, we use the difference between the 30-hour
accumulated and the 6-hour accumulated precipitation forecast.  For
initialization at 12 UTC, we use the difference between the 42-hour
accumulated and the 18-hour accumulated precipitation forecast, and
for longer accumulation times, we extend correspondingly.

To obtain forecasts for a specific station location from gridded NWP
forecasts, bilinear interpolation as well as a nearest neighbor
approach are possible.  We use the latter, implying that the forecast
for the station is the same as the forecast for the grid cell
containing the station.  Especially for large gridbox sizes, bilinear
interpolation may not be physically persuasive, and the nearest
neighbor approach is more compelling.

TRMM observations are temporally aggregated to the same periods as the
station observations. As they do not cover the exactly same periods, 
the first and last 3-hour TRMM observations are weighted by 0.5. For 
evaluation on different spatial scales, NWP forecasts and TRMM 
observations are aggregated to longitude--latitude
boxes of $0.25^\circ \times 0.25^\circ$, $1^\circ \times 1^\circ$, and
$5^\circ \times 2^\circ$.  As propagation of
precipitation systems is a potential error source and in an
environment with predominantly westward movement of them, the
largest box is tailored to assess NWP forecast quality without this
potential source of error.

\subsection{Consistency between TRMM and station observations}

\begin{figure}[t]
\centering
\includegraphics[width = 0.95 \textwidth]{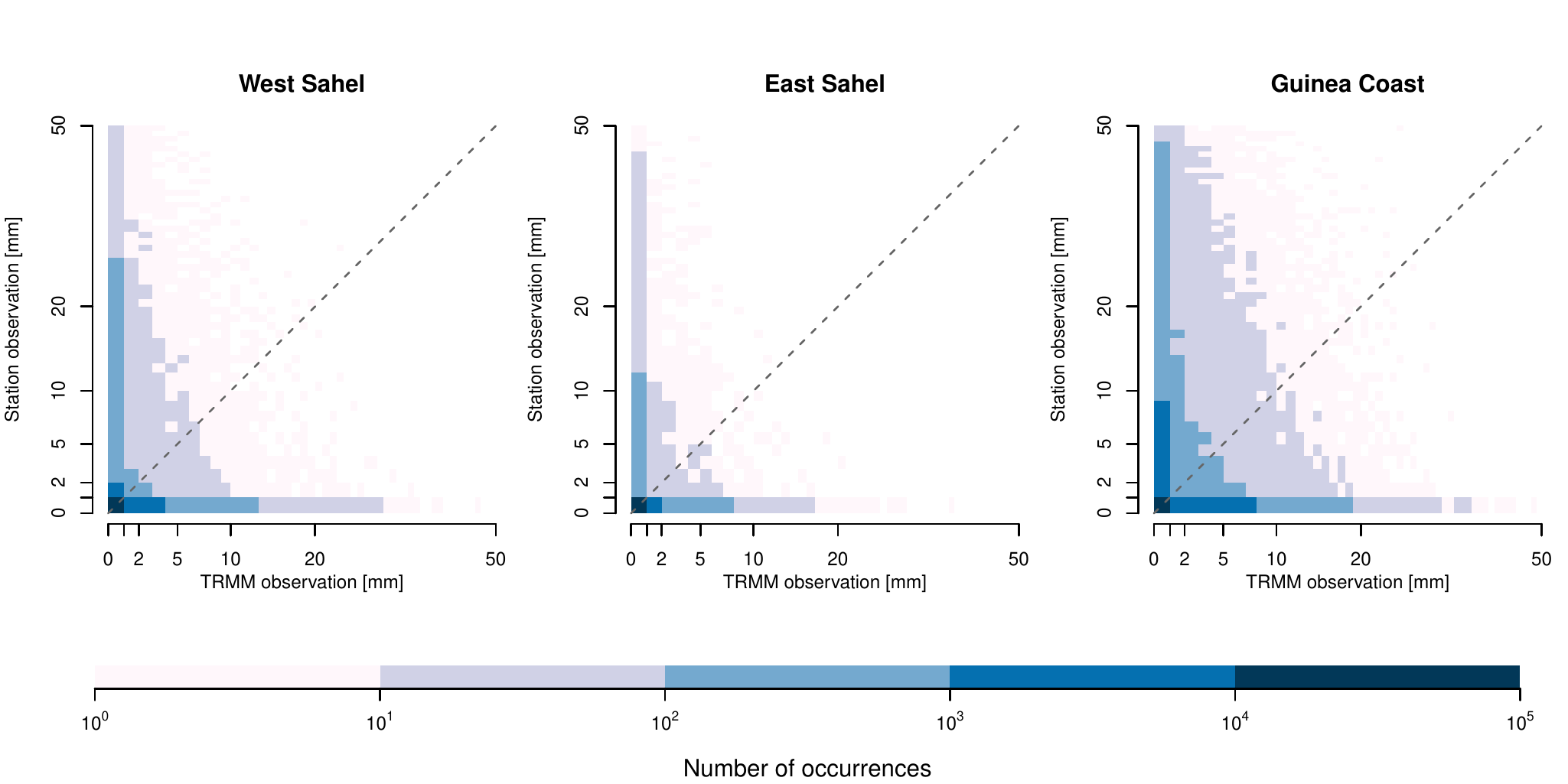}
\caption{Comparison of 1-day accumulated station and TRMM observations
  of precipitation in monsoon seasons 2007--2014.  Observations above
  50 mm exist, but are very infrequent.  \label{fig:TRMM}}
\end{figure}

In light of the dry bias of TRMM observations, we evaluate the
consistency of TRMM and station observations in our data sets.  Figure
\ref{fig:TRMM} displays a two-dimensional histogram for TRMM against
station observations based on all eight monsoon seasons for each of
the three regions.  In case TRMM and station observations agree, they
are close to the diagonal.  Below (above) the diagonal the station
observation is lower (higher) than the TRMM observation.  High spatial
variability of precipitation, different spatial coverage (point
vs.~about 625 km$^2$), and retrieval problems can lead to such
discrepancies.  Overall, the agreement between station and TRMM
observations is fair, and on average, station observations are
slightly higher than TRMM observations, consistent with the
literature.

\section{Methods}  \label{sec:Methods}  

Probabilistic forecasts are meant to provide calibrated information
about future events.  To be of use, they should satisfy two
properties.  First, they should convey correct probabilistic
statements, in that observations behave like random draws from the
forecast distributions.  This property is called calibration.  Second,
under all calibrated forecasts, sharper ones with lesser uncertainty
are preferred.  

\subsection{Reference forecasts}  \label{subsec:Reference}

For the assessment of raw and postprocessed ensemble forecast skill,
the availability of a benchmark forecast is essential.  Here we
introduce the concept of a probabilistic climatology that consists of
the observations during the 30 years prior to the considered year at
the considered day of the year and location.  This can be understood
as a 30-member observation-based ensemble forecast.  We extend the
probabilistic climatology by including observations in a $\pm$ 2-day
range around the considered day, and refer to this as Extended
Probabilistic Climatology (EPC).  

\cite{Hamill2006} note that pooling can lead to a deterioration when
performed across data with differing climatologies, leading to a
perceived, but incorrect improvement of assessed model forecast skill.
In our case, however, neighboring daily climatologies are very similar
and the pooling is performed over a range of $\pm$ 2 days only.  EPC
has better forecast quality than standard probabilistic climatology
(not shown) and is used as benchmark in the following.  As TRMM
observations are available for the period 1998--2014 only, the
TRMM-based EPC relies on this period without the considered
verification year.

\subsection{Assessing calibration: Unified probability integral transform (uPIT) histograms}  
\label{subsec:Calibration}

Verification rank histograms and probability integral transform (PIT)
histograms are standard tools for the assessment of calibration, and
we refer the reader to \citet{Hamill2001}, \citet{Gneiting2007} and
\citet{Wilks2011} for in-depth discussions of their use and
interpretation.  In a nutshell, for calibrated probabilistic
forecasts, rank and PIT histograms are uniform, U-shaped histograms
indicate underdispersion, and skewed histograms mark biases.

For an ensemble forecast, the verification rank is the rank of the
observation when it is pooled with the $m$ ensemble members; clearly,
this is an integer between 1 and $m + 1$.  If $k$ members predict no
precipitation, and no precipitation is observed, the rank is randomly
drawn between 1 and $k + 1$.  For a probabilistic forecast in the form
of a cumulative distribution function (CDF), $F$, and a verifying
precipitation accumulation $y > 0$, the PIT is the value $F(y)$ of the
forecast CDF evaluated at the observation.  In the case of no
precipitation, a value is randomly drawn between 0 and the forecast
probability of no precipitation \citep{Sloughter2007}.

In the present study, we compare raw ensemble forecasts to
postprocessed forecasts in the form of CDFs, and the TIGGE
sub-ensembles have varying numbers of members.  We use the term
probabilistic quantitative precipitation forecast (PQPF) to denote all
these types of forecasts.  To allow a compelling visual assessment of
calibration in this setting, we introduce the notion of a unified PIT
(uPIT).  For a forecast in the form of a CDF, the uPIT is simply the
PIT.  For an ensemble forecast with $m$ members, if the observation
has rank $i$ and this rank is unique, the uPIT is a random number from
a uniform distribution between $\frac{i-1}{m + 1}$ and $\frac{i}{m +
1}$.  If $k$ members predict no precipitation, and no precipitation
is observed, the uPIT is a random number between 0 and $\frac{k+1}{m +
1}$.  It is readily seen that for a calibrated PQPF the uPIT is
uniformly distributed.  Hereinafter, we use 20 equally spaced bins to
plot uPIT histograms, and we simply talk of PIT values.

Our uPIT histograms focus on calibration regarding the forecasted
precipitation amount.  However, any PQPF induces a probability of
precipitation (PoP) forecast for the binary event of rainfall
occurrence at any given threshold value.  We use a threshold of 0.2 mm
to define rainfall occurrence, but the results reported on hereinafter
are insensitive to this choice.  Reliability, the equivalent of
calibration for probability forecasts of binary events, means that
events declared to have probability $p$ occur a proportion $p$ of the
time.  This can be checked empirically in reliability diagrams, where
the observed frequency of occurrence is plotted vs.~the forecast
probability \citep[e.g.,][]{Wilks2011}.

\subsection{Proper scoring rules}

For the comparative evaluation of predictive skill we use proper
scoring rules that assess calibration and sharpness simultaneously
\citep{Gneiting2007a, Wilks2011}.  Specifically, the Continous Ranked
Probability Score (CRPS) for a PQPF with CDF $F$ and a verifying
observation $y$ is defined as
 \begin{equation*}  
 \text{CRPS}(F,y) =  \int_{-\infty}^\infty [F(x) - \myI (x \geq y) ]^2 \: {\rm d}x, 
 \end{equation*}
where $\myI$ is an indicator function, equal to 1 if the argument is
true and equal to 0 otherwise.  From every PQPF, we can extract a
deterministic forecast and compute its absolute error (AE).  If the
deterministic forecast is chosen to be the median of the forecast
distribution, the AE can be interpreted as a proper scoring rule
\citep{Gneiting2011, Pinson2012}.  Both the AE and the CRPS are
negatively oriented, and they are reported in the unit of the
observation (here, millimeter), and so can be compared directly.  In
fact, if the forecast distribution is a deterministic forecast, the
CRPS reduces to the AE \citep{Gneiting2007a}.

With the probability of precipitation (PoP) being an essential
component of a PQPF, the evaluation of PoP forecast quality by proper
scoring rules is desirable, and can be accomplished by means of the
Brier score \citep[BS,][]{Brier1950}.  For a probability forecast $p$
for a binary event to occur, the negatively oriented BS is $(1-p)^2$
if the event occurs and $p^2$ if it does not occur.

It is well known that not only the BS, but many proper scoring rules
for probability forecasts of binary events exist, and that forecast
rankings can depend on the choice of the proper scoring rule.
However, every proper scoring rule admits a representation as a
weighted average over so-called elementary scores, $\text{S}_\theta$,
which can be interpreted economically, with the parameter $\theta$
representing a decision maker's cost--loss ratio.  \cite{Ehm2016}
advocate the use of so-called Murphy diagrams, which display, for each
forecast considered, the mean elementary score as a function of the
cost-loss ratio $\theta \in (0,1)$.  If a forecast receives lower
elementary scores than another for all $\theta$, then it is preferable
for any decision maker, and receives lower scores under just any
proper scoring rule.

Interestingly, the area under a forecast's graph in a Murphy diagram
equals half its mean BS \citep{Ehm2016}.

\subsection{Statistical postprocessing}  \label{subsec:Postprocessing}

Statistical postprocessing addresses structural deficiencies of NWP
model output.  Here we use the well established methods of Ensemble
Model Output Statistics \citep[EMOS,][]{Gneiting2005, Scheuerer2014b}
and Bayesian Model Averaging \citep[BMA,][]{Raftery2005a,
 Sloughter2007} to correct for systematic errors in ensemble
forecasts of precipitation accumulation.  

In this subsection, we review these methods with focus on the
52-member ECMWF EPS, and we denote the values of its HRES, CNT, and
ENS members by $x_\HRES$, $x_\CNT$, and $x_1, \ldots, x_{50}$,
respectively.  We write $\bar{x}_\ENS$ for the mean of the ENS
members, $\bar{p}$ for the fraction (out) of (all 52) members that
predict no precipitation, and denote the observed precipitation
accumulation by $y$.  Adaptations of the postprocessing schemes to the
other TIGGE sub-ensembles and the reduced multi-model (RMM) ensemble
are straightforward.

\subsubsection{Ensemble Model Output Statistics (EMOS)}  \label{subsec:EMOS}

The idea of the EMOS approach is to convert an ensemble forecast into
a parametric distribution, based on the ensemble forecast at hand
\citep{Gneiting2005}.  \cite{Scheuerer2014b} introduced an EMOS
approach for precipitation accumulations that relies on the
three-parameter family of left-censored Generalized Extreme Value
(GEV) distributions.  The left-censoring allows for a point mass at
zero, and the shape parameter for flexible skewness in positive
precipitation accumulations.

Briefly, the EMOS predictive distribution based on the ECMWF ensemble
is a left-censored GEV distribution with location parameter that is an
affine function of $x_\HRES$, $x_\CNT$, $\bar{x}_\ENS$ and $\bar{p}$,
a scale parameter that is an affine function of the ensemble mean
difference, and a shape parameter that is estimated from training
data, but does not link to the ensemble values \citep{Scheuerer2014b}.

\begin{figure}[t]
\centering
\includegraphics[width = 0.79 \textwidth]{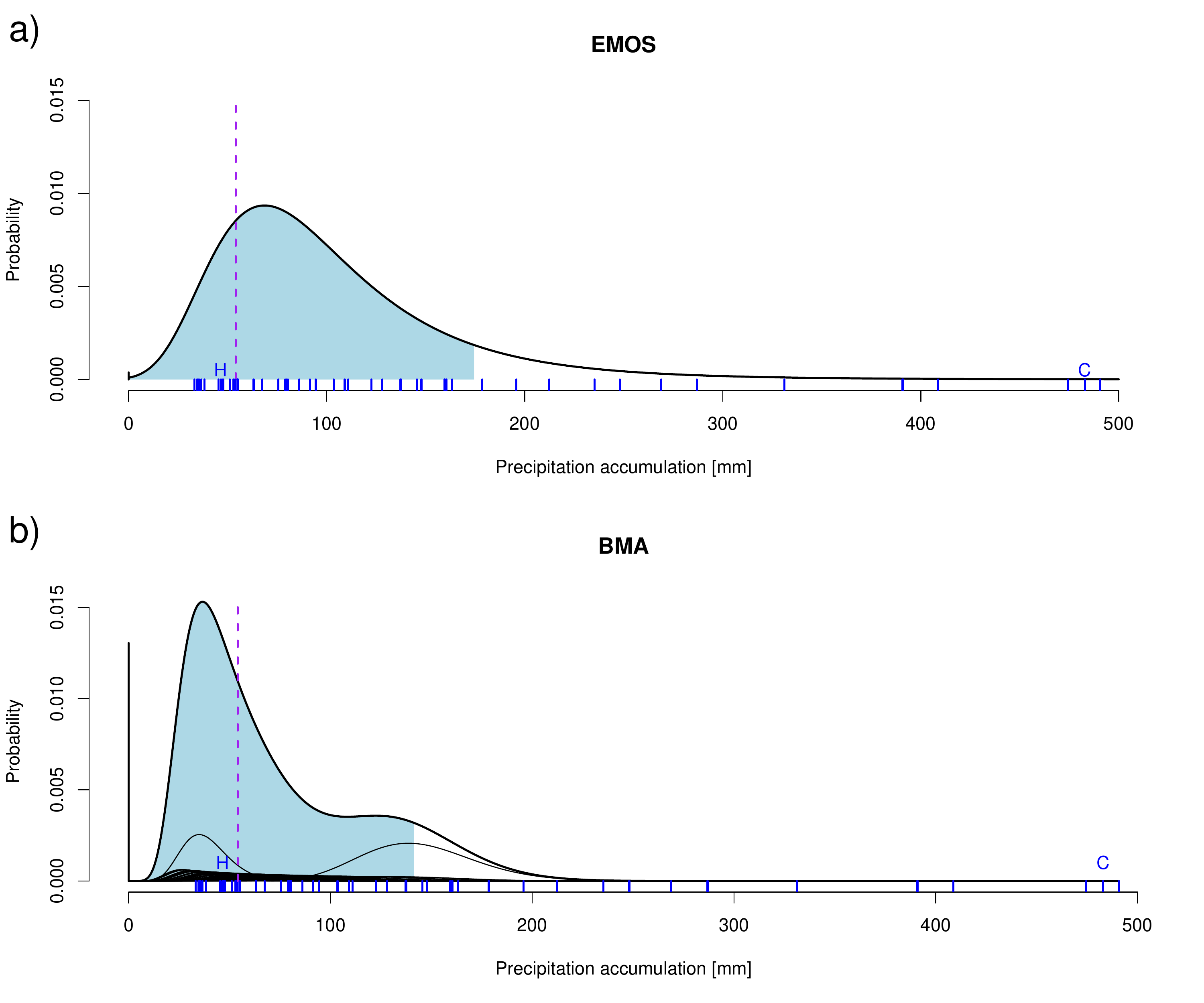}
\caption{EMOS and BMA postprocessed ECMWF ensemble forecasts for 5-day
  accumulated precipitation at Ougadougou, Burkina Faso, valid 03 Aug
  -- 08 Aug 2007.  The blue marks at bottom represent the 52 raw ECMWF
  ensemble members, including the HRES (H) run, the CNT (C) run, and
  the 50 perturbed ENS members.  (a) The EMOS postprocessed forecast
  includes a tiny point mass at zero and a censored GEV density for
  positive accumulations.  (b) The BMA postprocessed forecast includes
  a point mass at zero, which is represented by the solid bar, and a
  mixture of power transformed Gamma densities for positive
  accumulations.  The 52 component densities are represented by the
  thin black curves, with the HRES and CNT components standing out.
  The lower 90\% prediction interval is indicated in light blue, and
  the dashed bar represents the verifying precipitation
  accumulation.  \label{fig:EMOS.BMA}}
\end{figure}

For illustration, Fig.~\ref{fig:EMOS.BMA}a shows an EMOS
postprocessed forecast distribution for 5-day accumulated
precipitation at Ougadougou, Burkina Faso.  The 52 raw ECMWF ensemble
members are represented by blue marks; they include eleven values in
excess of 200 mm, with the CNT member being close to 500 mm.  The
ensemble forecast at hand informs the statistical parameters of the
EMOS postprocessed forecast distribution, which includes a tiny point
mass at zero, and a censored GEV density for positive precipitation
accumulations, with the 90th percentile being at 174 mm.

\subsubsection{Bayesian Model Averaging (BMA)}  \label{subsec:BMA}

A BMA predictive distribution is a weighted sum of component
distributions, each of which depends on a single ensemble member.  For
the ECMWF ensemble, the BMA method for precipitation accumulation
proposed and studied by \cite{Sloughter2007} and \cite{Fraley2010}
implies a statistical model of the form
\begin{eqnarray*}  
y \, | \, x_\HRES, x_\CNT, x_1, \ldots, x_{50}  \sim w_\HRES \, g_\HRES(y|x_\HRES) \\
+ \: w_\CNT \, g_\CNT(y|x_\CNT)  + \frac{w_\ENS}{50} \, \sum_{i=1}^{50} g_\ENS(y|x_i), 
\end{eqnarray*}
with nonnegative weights $w_\HRES$, $w_\CNT$, and $w_\ENS$ that sum to
1, and reflect the members' performance in the training period.  The
component distributions $g_\HRES$, $g_\CNT$, and $g_\ENS$ include a
point mass at zero and a density for positive accumulations.  The
point mass at zero specifies the probability of no precipitation in a
logistic regression model, with the cube root of the member forecast
and an indicator of it being zero, being linear predictors.  The
specification for positive amounts derives from a Gamma density for
the cube root transformed precipitation accumulation, with a mean that
is affine in the cube root transformed ensemble member, and a variance
that is an affine function of the member value.  While the statistical
coefficients for the Gamma mean model are estimated for $g_\HRES$,
$g_\CNT$, and $g_\ENS$ separately, the coefficients for the Gamma
variance model are shared.

Figure \ref{fig:EMOS.BMA}b shows a BMA postprocessed
forecast distribution for the aforementioned forecast case at
Ougadougou, Burkina Faso.  The postprocessed distribution involves a
point mass of about 0.01 at zero, and a mixture of power transformed
Gamma densities for positive accumulations, with the 90th percentile
being at 141 mm.  In this example, the BMA and EMOS postprocessed
distributions are sharper than the raw ECMWF ensemble, and
nevertheless the verifying accumulation is well captured.

Adaptations to the other ensembles considered in this paper are
straightforward as described by \cite{Fraley2010}.  For example, in
the case of the RMM ensemble each of the 15 contributors receive its
own component distribution, BMA weight, logistic regression
coefficients for the probability of no precipitation, and statistical
parameters for the Gamma mean model, whereas the coefficients for the
Gamma variance model are shared.

\subsubsection{Estimation of statistical parameters}  \label{subsec:Estimation}

Postprocessing techniques such as EMOS and BMA rely on statistical
parameters that need to be estimated from training data, comprising
forecast--observation-pairs either from the station or TRMM pixel at
hand, or from all stations or applicable TRMM pixels within the
considered region, and typically from a rolling training period
consisting of the $n$ most recent days for which data are available at
the initialization time.  We employ the regional approach with a
rolling training period of $n = 20$ days which yields
superior results, consistent with the literature
\citep[e.g., ][]{Thorarinsdottir2010}. The local approach requires
longer training periods, and (in experiments not shown here) yields
very similar results then.

For EMOS, parameter estimation is based on CRPS minimization over the
training data, which is computationally efficient, as closed
expressions for the CRPS under GEV distributions are available
\citep{Scheuerer2014b}.  For BMA, we employ maximum likelihood
estimation, implemented via the expectation-maximization (EM)
algorithm developed by \citet{Sloughter2007}.  All computations were
performed in R \citep{R} based on the {\tt ensembleBMA} package
\citep{Fraley2011} and code supplied by Michael Scheuerer.

\section{Results}  \label{sec:Results}

Our annual evaluation period ranges from 1 May to 15 October, covering
the wet period of the West African monsoon.  The assessment of ECMWF
ensemble forecasts is based on monsoon seasons 2007--2014, and for the
other TIGGE sub-ensembles we restrict according to availability as
indicated in Table \ref{tab:TIGGE}.

For verification against station observations, this yields more than
3,000, 6,000, and 12,000 forecast--observations pairs per monsoon
season in East Sahel, West Sahel, and Guinea Coast.  For verification
against TRMM observations, we use 30 randomly chosen, non-overlapping
boxes per region at $0.25^\circ \times 0.25^\circ$ and $1^\circ \times
1^\circ$ aggregation, and eight sites per region for $5^\circ \times
2^\circ$ longitude--latitude boxes.  This covers substantial parts of
the study region and results in about 5,000 forecast--observation
pairs per monsoon season at the smaller aggregation levels, and well
over 1,000 pairs at our highest level.

In subsection \ref{subsec:ECMWF24}, we study the
skill of 1-day accumulated ECMWF raw and postprocessed ensemble
precipitation forecasts in detail.  Subsections
\ref{subsec:ECMWF120} and \ref{subsec:TRMM} present
results and highlight differences for longer accumulation times and
spatially aggregated forecasts.  Subsection
\ref{subsec:TIGGE} turns to results for all TIGGE
sub-ensembles, and we investigate the gain in predictability through
inter-model variability using the RMM ensemble.  In our (u)PIT
histograms and reliability diagrams, we show results for the last
available monsoon season 2014 only, given that operational systems continue
to be improving \citep{Hemri2014}.

\subsection{1-day accumulated ECMWF forecasts}  \label{subsec:ECMWF24}

\begin{figure}[pt]
\centering 
\includegraphics[width = 0.8\textwidth]{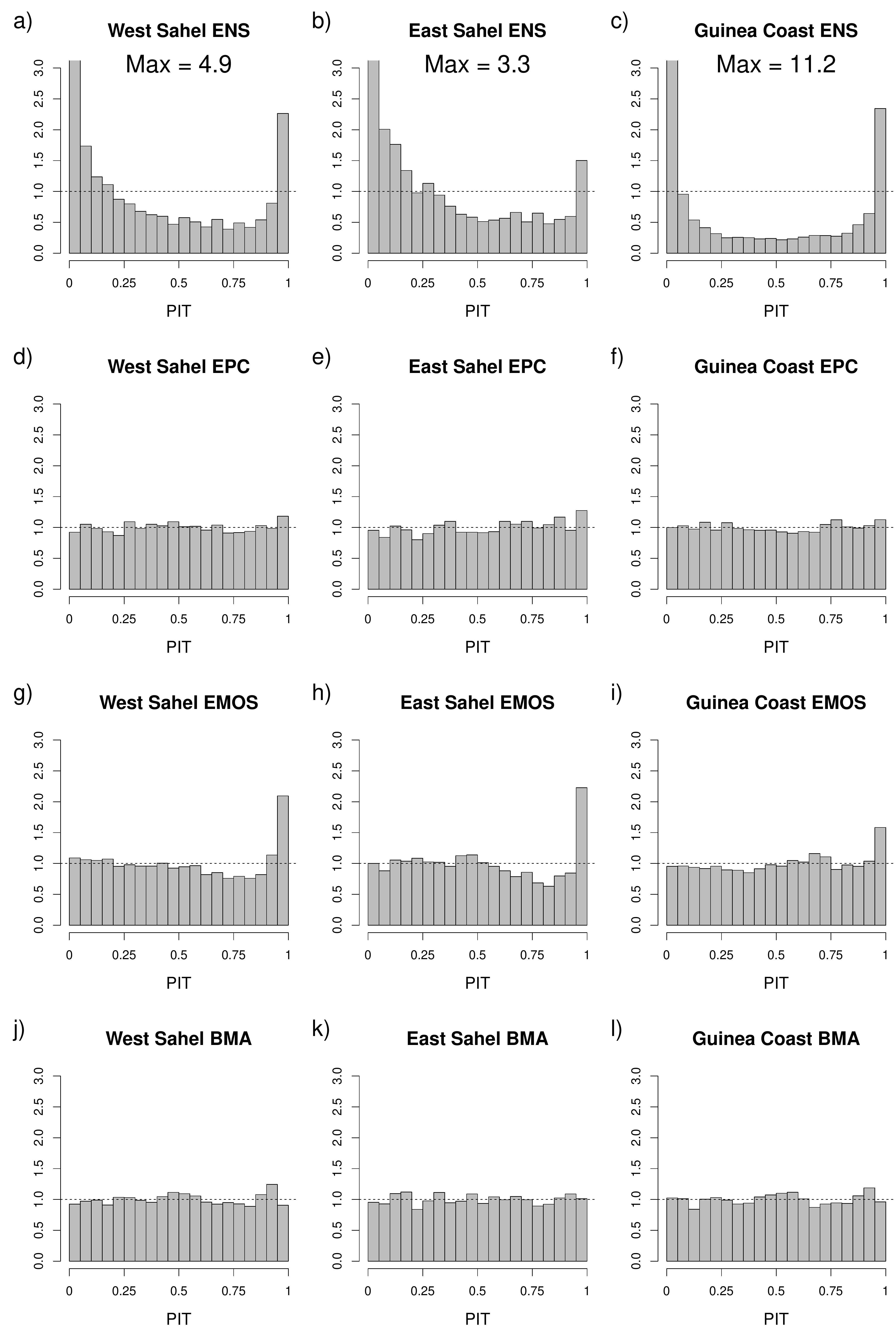}
\caption{Unified PIT (uPIT) histograms for raw ECMWF ensemble, EPC,
  and EMOS and BMA postprocessed forecasts of 1-day accumulated
  precipitation in monsoon season 2014, verified against station
  observations.  Histograms are cut at a height of 3, with the
  respective maximal height noted.  The dashed line indicates the
  uniform distribution that corresponds to a calibrated
  forecast.  \label{fig:ECMWF24.PIT}}
\end{figure}

\begin{figure}[pt]
\centering
\includegraphics[width = 0.80 \textwidth]{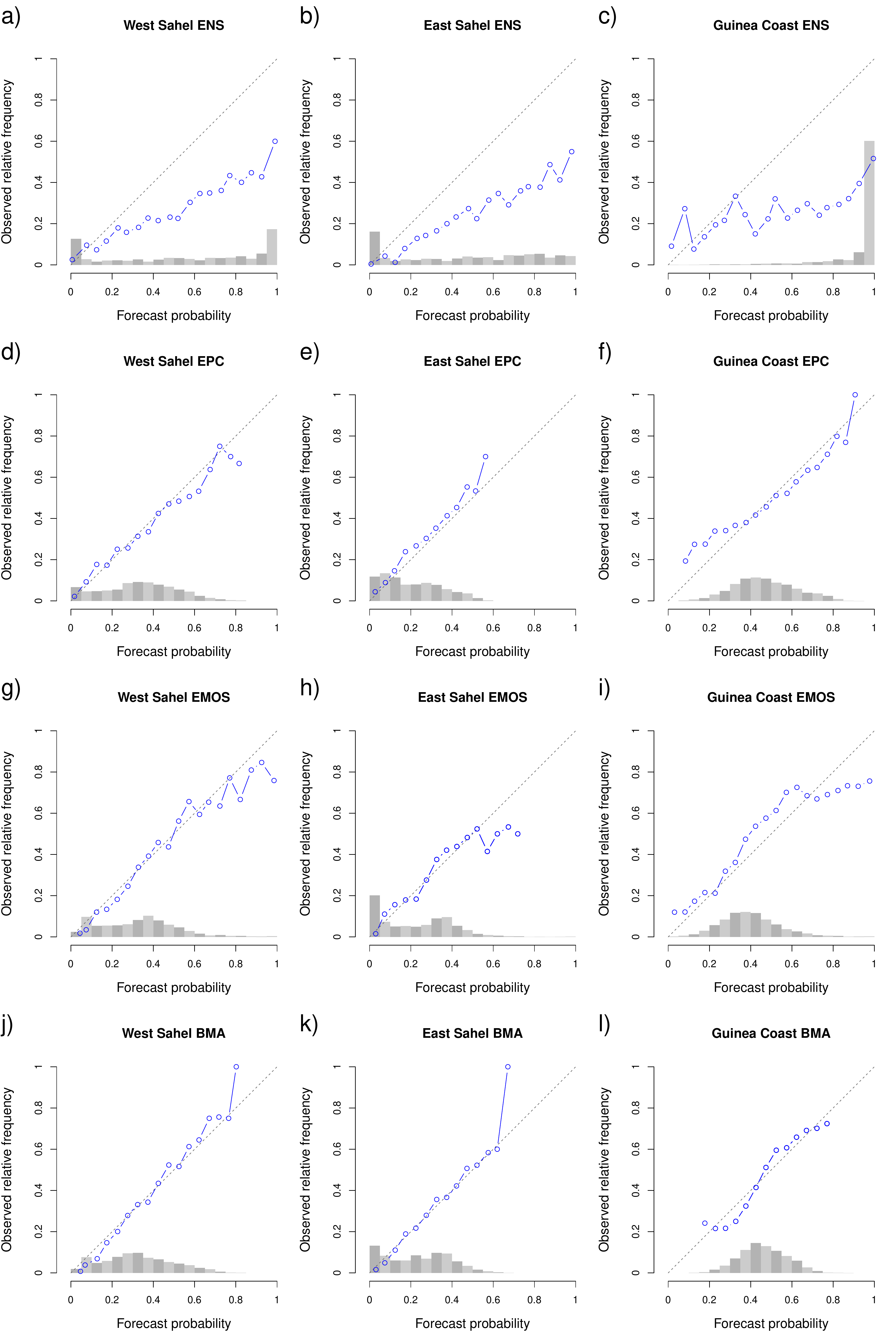}
\caption{Reliability diagrams for raw ECMWF ensemble, EPC, and EMOS
  and BMA postprocessed forecasts of 1-day accumulated precipitation
  in monsoon season 2014, verified against station observations.  The
  diagonal indicates perfect reliability, and the histograms show the
  relative frequencies of the PoP forecast
  values.  \label{fig:ECMWF24.Reliability}}
\end{figure}

Figure \ref{fig:ECMWF24.PIT} shows (u)PIT histograms for 1-day
accumulated raw and postprocessed ECMWF ensemble and EPC forecasts
over West Sahel, East Sahel, and Guinea Coast.  The histograms for the
raw ensemble indicate strong underdispersion as well as a wet bias
(panels a--c).  At Guinea Coast, about 56\% of the
observations are smaller than the smallest ensemble member, a result
that is robust across monsoon seasons.  EMOS and BMA postprocessed
forecasts generally are calibrated (panels g--l), as
is EPC (panels d--f).  Statistical postprocessing
also corrects for the systematically too high PoP values issued by the
raw ECMWF ensemble.  As shown in Fig.~\ref{fig:ECMWF24.Reliability},
EMOS and BMA postprocessed PoP forecasts are reliable, but are hardly
ever higher than 0.70.  Generally, the postprocessed PoP forecasts
have reliability and resolution similar to EPC.

Table \ref{tab:ECMWF24} shows the mean Brier score (BS), mean CRPS,
and mean absolute error (MAE) for the various forecasts and regions,
with the scores being averaged across monsoon seasons 2007--2014.  We
use a simple procedure to check whether differences in skill are
stable across seasons.  If a method has higher (worse) mean score than
EPC in all eight seasons, we mark the score \ww; if it is judged worse
in seven seasons, we put down a \wc.  Similarly, if a method has
smaller (better) mean score than EPC in all seasons, we mark the score
\bb; if it performs better in seven seasons, we label \bc \ in the
table.  Viewed as a (one-sided) statistical test of the hypothesis of
predictive skill equal to EPC, the associated tail probabilities or
$p$-values are $1/2^8 = 0.0039\ldots$ and $(1 + 8)/2^8 = 0.035\ldots$
respectively.  Clearly, the raw ECMWF ensemble underperforms relative
to EPC, with \ww \ designations throughout, and the EMOS and BMA
postprocessed forecasts perform at about the same level as EPC.

\begin{table}[t]
\caption{Mean Brier score (BS), mean CRPS, and MAE for raw ECMWF
  ensemble, EPC, and EMOS and BMA postprocessed forecasts of 1-day
  accumulated precipitation in monsoon seasons 2007--2014, verified
  against station observations.  If a method has a higher (worse)
  respectively lower (better) mean score than EPC in 
  all eight seasons, the score is marked \ww\ respectively \bb; 
  if it performs worse respectively better than EPC in seven seasons, 
  the score is marked \wc \ respectively \bc.  \label{tab:ECMWF24}}
\medskip
\centering
\resizebox{\textwidth}{!}{
\begin{tabular}{l|rrr|rrr|rrr}
\hline
& \multicolumn{3}{c}{BS} &  \multicolumn{3}{c}{CRPS} &  \multicolumn{3}{c}{MAE}  \\ \vspace{-3mm}
& West \hspace{0.02mm} & East $\,$ & Guinea & West \hspace{0.02mm} & East $\,$ & Guinea & West \hspace{0.02mm} & East $\,$ & Guinea \\[5pt]
& Sahel & Sahel & Coast $\,$ & Sahel & Sahel & Coast $\,$ & Sahel & Sahel & Coast $\,$ \\
\hline
ENS  & \ww0.32 & \ww0.32 & \ww0.48 & \ww4.50 & \ww2.63 & \ww6.99 & \ww5.36 & \ww3.13 & \ww8.39 \\ 
EPC  &    0.19 &    0.15 &    0.23 &    3.75 &    2.08 &    5.28 &    4.60 &    2.38 &    6.57 \\ 
EMOS &    0.19 &    0.15 &    0.23 &    3.75 &    2.15 & \bc5.25 &    4.65 & \ww2.45 &    6.60 \\ 
BMA  & \bc0.18 &    0.15 & \bb0.22 &    3.71 &    2.07 & \bb5.20 &    4.58 &    2.38 &    6.53 \\ 
\hline
\end{tabular}
}
\end{table}

\begin{figure}[t]
\centering
\includegraphics[width = \textwidth]{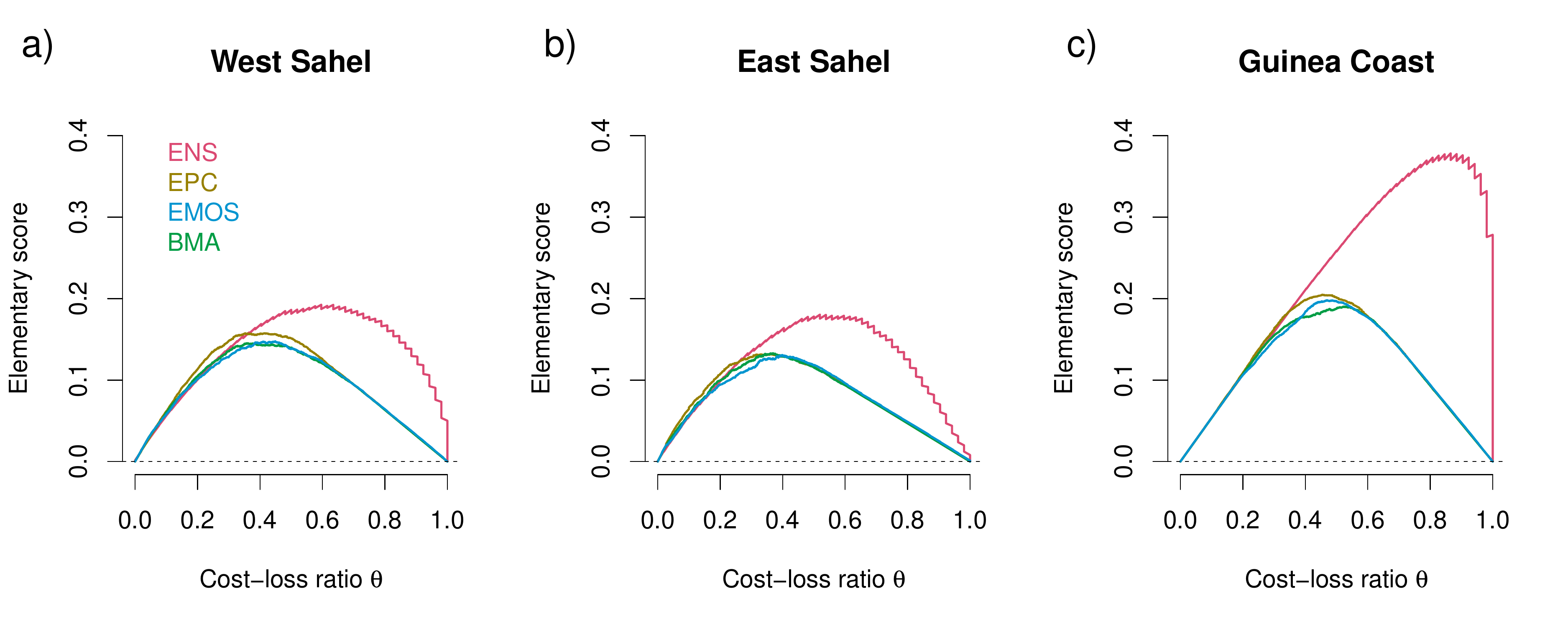}
\caption{Murphy diagrams for raw ECMWF ensemble (ENS), EPC, and EMOS
  and BMA postprocessed 1-day accumulated PoP forecasts in monsoon
  season 2014, verified against station observations.  
\label{fig:Murphy}}
\end{figure}

The Murphy diagrams in Fig.~\ref{fig:Murphy} corroborate these
findings.  For 1-day precipitation occurrence, decision makers will
mostly prefer the climatological reference EPC over the raw ECMWF
ensemble, and only some decision makers will have a slight preference
for EMOS or BMA postprocessed forecasts, as compared to EPC.  These
are sobering results, as they suggest that over northern tropical
Africa ECMWF 1-day accumulated precipitation forecasts are hardly of
use.

\subsection{Longer accumulation times}  \label{subsec:ECMWF120}

One might expect NWP precipitation forecasts to improve relative to
EPC at longer accumulation times, as the main focus in forecasting
shifts from determining time and location of initiation and subsequent
propagation of convection towards determining regions with enhanced or
reduced activity, based on large-scale conditions.  Longer lead times
might also lead to growth in differences between perturbed members,
and thus reduce raw ensemble underdispersion.  

The PIT histogram in of Fig.~\ref{fig:ECMWF120}a indicates
only slight, if any, improvement in calibration for raw ECMWF 5-day
accumulated precipitation forecasts over West Sahel, and the results
for the other regions are similar (not shown).  Raw ensemble
reliability improves at longer accumulation times, verified against
either station observations in panel b), or $5^\circ \times 2^\circ$
TRMM observations in panels c) and d), though at a loss of resolution.

\begin{table}[t]
\caption{Mean Brier score (BS), mean CRPS, and MAE for raw ECMWF
 ensemble, EPC, and EMOS and BMA postprocessed forecasts of 5-day
 accumulated precipitation in monsoon seasons 2007--2014, verified
 against station observations.  Same setup as in
 Table~\ref{tab:ECMWF24}.  \label{tab:ECMWF120}}
\medskip
\centering
\resizebox{\textwidth}{!}{
\begin{tabular}{l|rrr|rrr|rrr}
\hline
& \multicolumn{3}{c}{BS} &  \multicolumn{3}{c}{CRPS} &  \multicolumn{3}{c}{MAE}  \\ \vspace{-3mm}
& West \hspace{0.02mm} & East $\,$ & Guinea & West \hspace{0.02mm} & East $\,$ & Guinea & West \hspace{0.02mm} & East $\,$ & Guinea \\[5pt] 
& Sahel & Sahel & Coast $\,$ & Sahel & Sahel & Coast $\,$ & Sahel & Sahel & Coast $\,$ \\
\hline
ENS  &    0.14 & \ww0.25 & \ww0.10 & \ww12.80 & \ww8.42 & \ww19.69 &    16.23 & \ww10.76 & \wc24.41 \\ 
EPC  &    0.12 &    0.16 &    0.08 &    11.63 &    7.07 &    16.54 &    16.15 &     9.56 &    22.98 \\ 
EMOS &    0.13 &    0.16 & \wc0.08 &    11.62 &    7.34 &    16.44 & \bc15.99 &     9.96 &    22.74 \\ 
BMA  & \bc0.11 & \bb0.15 &    0.08 & \bb11.47 & \bc6.94 &    16.33 & \bc16.07 &  \bc9.45 &    22.92 \\ 
\hline
\end{tabular}
}
\end{table}

Table \ref{tab:ECMWF120} uses the same setting as Table
\ref{tab:ECMWF24}, but the scores are now for 5-day accumulated
precipitation.  The raw ECMWF ensemble still underperforms relative to
EPC.  The EMOS and BMA postprocessed forecasts outperform EPC only
slightly, with the differences in scores being small and typically not
being stable across monsoon seasons.  Despite the change in the
underlying forecast problem, even postprocessed ECMWF ensemble
forecasts are generally not superior to EPC.

\subsection{Spatially aggregated observations}  \label{subsec:TRMM}

For the assessment of forecast skill at larger spatial scales, we
focus on ECMWF raw and BMA postprocessed ensemble forecasts over West
Sahel, evaluated by the Brier score and CRPS.  This is due to the
similarities in CRPS and MAE results, better performance of 

\begin{figure}[tp]
\centering \includegraphics[width = 0.5\textwidth]{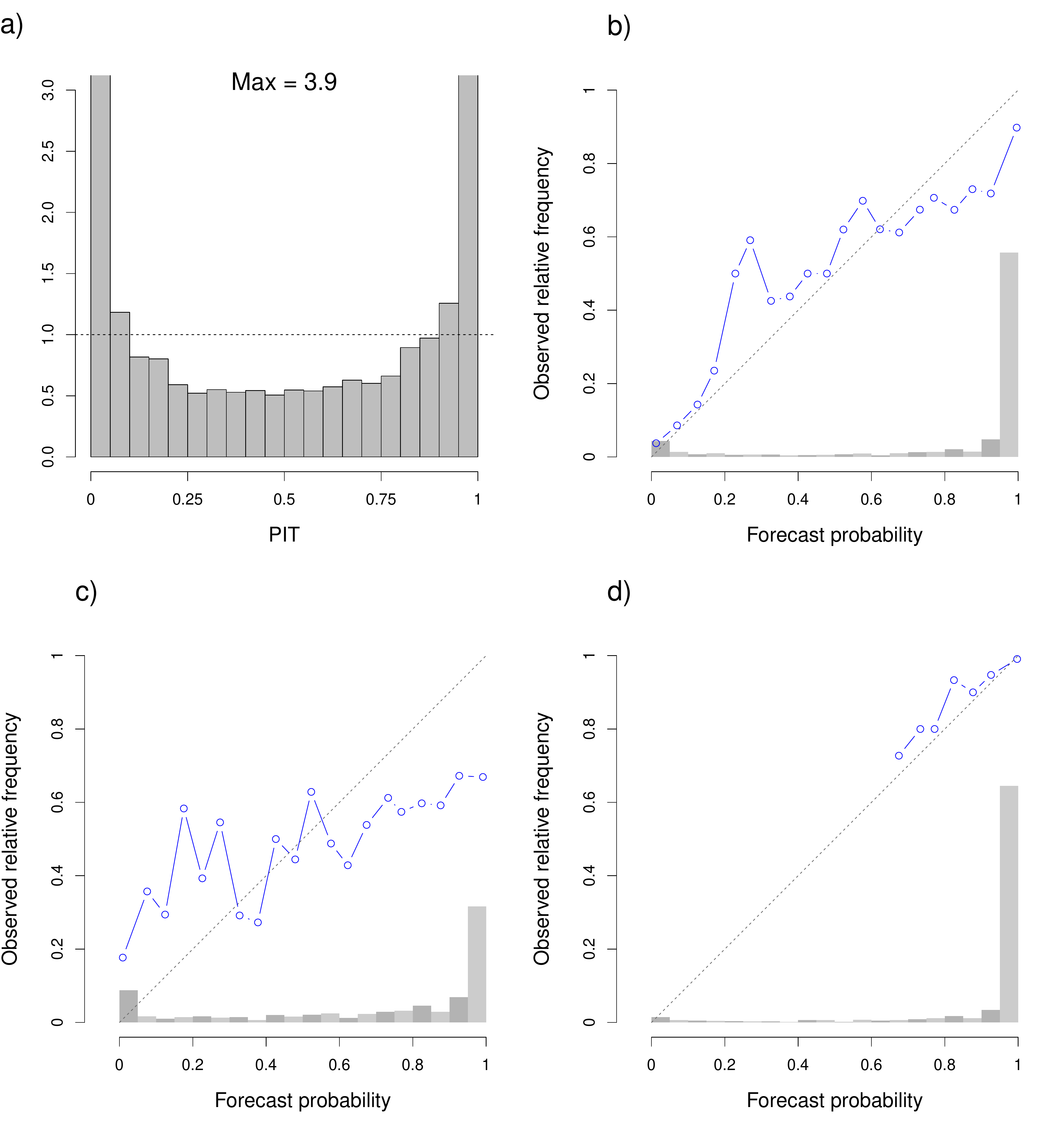}
\caption{Calibration and reliability of raw ECMWF ensemble forecasts
 over West Sahel in monsoon season 2014 at 1- and 5-day
 accumulations.  a) (u)PIT histogram and b) reliability diagram for
 5-day accumulated precipitation, verified against station
 observations.  Panels c) and d) show reliability diagrams for 1-
 and 5-day accumulated precipitation, verified again $5^\circ \times
 2^\circ$ aggregated TRMM observations.  Same setup as in
 Figs.~\ref{fig:ECMWF24.PIT} and \ref{fig:ECMWF24.Reliability}.
 \label{fig:ECMWF120}}
\end{figure}

\begin{table}[bth]
\caption{Performance of spatially aggregated raw ECMWF ensemble, EPC,
 and BMA postprocessed forecasts of 1-day accumulated precipitation
 in monsoon seasons 2007--2014, verified against TRMM gridbox
 observations.  Same setup as in
 Table~\ref{tab:ECMWF24}.  \label{tab:TRMM}}
\medskip
\centering
\resizebox{\textwidth}{!}{
\begin{tabular}{l|rrr|rrr|rrr|rrr}
\hline
& \multicolumn{6}{c}{TRMM 0.25$^\circ \times$ 0.25$^\circ$ / 1 d} &  \multicolumn{3}{c}{TRMM 1$^\circ \times$ 1$^\circ$ / 1 d} 
& \multicolumn{3}{c}{TRMM 5$^\circ \times$2$^\circ$ / 1 d} \\
& \multicolumn{3}{c}{BS} &  \multicolumn{3}{c}{CRPS}  & \multicolumn{3}{c}{CRPS} &  \multicolumn{3}{c}{CRPS} \\ \vspace{-3mm}
& West \hspace{0.02mm} & East $\,$ & Guinea & West \hspace{0.02mm} & East $\,$ & Guinea 
& West \hspace{0.02mm} & East $\,$ & Guinea & West \hspace{0.02mm} & East $\,$ & Guinea \\[5pt]
& Sahel & Sahel & Coast $\,$ & Sahel & Sahel & Coast $\,$ & Sahel & Sahel & Coast $\,$ & Sahel & Sahel & Coast $\,$ \\
\hline
ENS & \ww0.30 & \ww0.23 & \ww0.48 & \ww2.29 & \ww1.44 & \ww4.03 & \ww2.24 & \ww1.56 & \ww4.43 & \ww1.95 & \ww1.53 & \ww4.22 \\ 
EPC &    0.19 &    0.14 &    0.23 &    1.07 &    0.57 &    1.35 &    0.94 &    0.58 &    1.36 &    0.81 &    0.49 &    1.07 \\  
BMA & \bb0.17 & \bb0.13 & \bb0.21 & \bb1.03 & \bb0.55 & \bb1.29 & \bb0.89 & \bb0.55 & \bb1.28 & \bb0.76 & \bb0.45 & \bb0.95 \\ 
\hline
\end{tabular}
}
\end{table}

BMA compared to EMOS in many instances, and results for West Sahel that
are as good for BMA postprocessed forecasts as for East Sahel, and
better than for Guinea Coast.

The use of spatially aggregated TRMM observations avoids problems of
point to pixel comparisons, and at higher aggregation we can assess
forecast quality with minimal propagation error.  The dry bias of TRMM
disadvantages the raw ensemble compared to EPC and postprocessed
forecasts, but does not hinder assessments regarding systematic
forecast errors.  As illustrated in Fig.~\ref{fig:ECMWF120}c, 1-day PoP
forecasts from the raw ECMWF ensemble remain unreliable even at the
$5^\circ \times 2^\circ$ gridbox scale. It is only under large scales
and longer accumulation times simultaneously, when precipitation
occurs almost invariably, that raw ensemble PoP forecasts become
reliable (panel d).

Table \ref{tab:TRMM} shows mean Brier and CRPS scores at various
spatial aggregations for 1-day precipitation accumulation, verified
against TRMM observations.  The raw ECMWF ensemble forecast is
inferior to EPC at all resolutions, and in every single region and
season.  BMA postprocessed forecasts outperform EPC across aggregation
scales, and in every single region and season, but the improvement
relative to EPC remains small.

\begin{figure}[pt]
\centering \includegraphics[width = \textwidth]{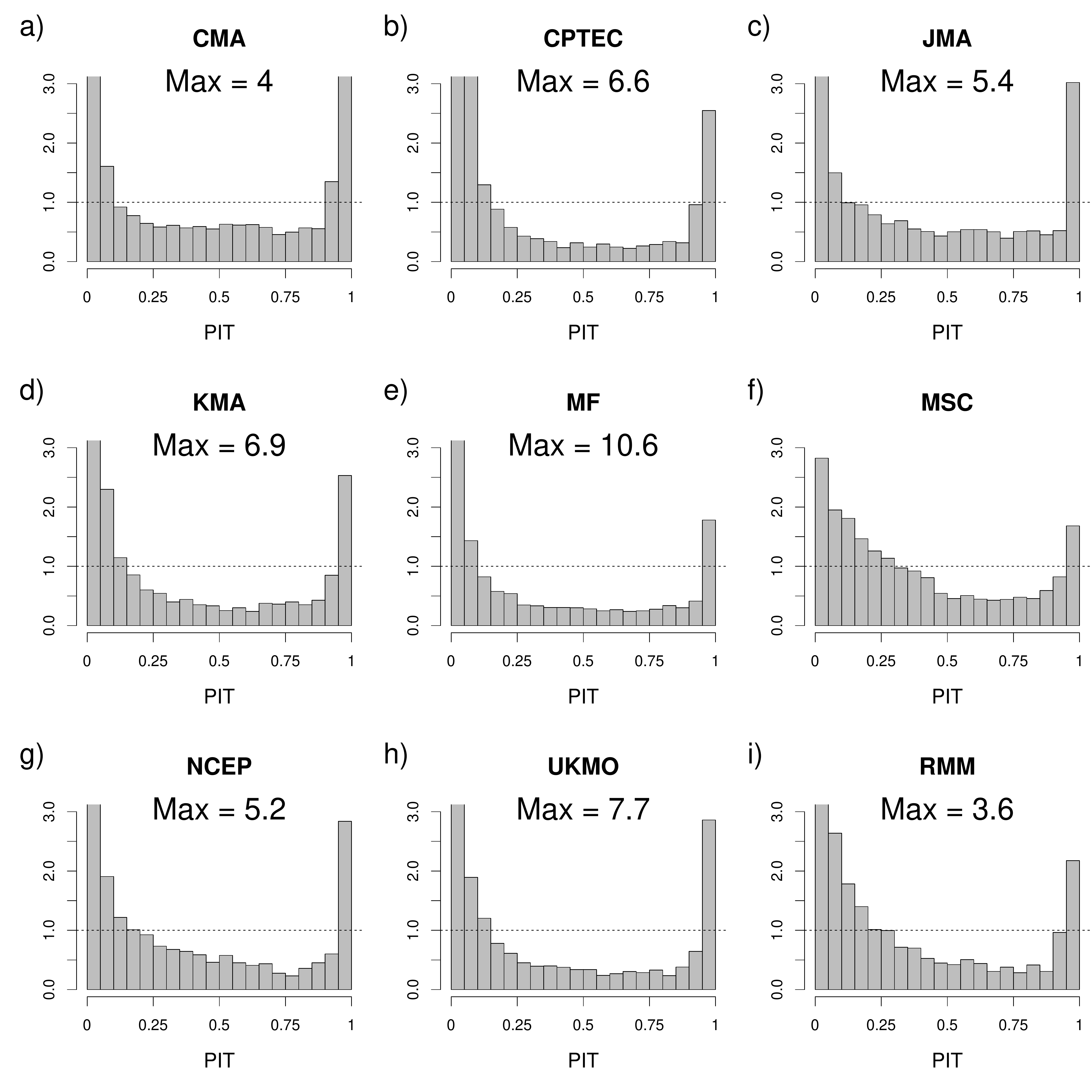}
\caption{(u)PIT histograms for raw TIGGE sub-ensemble and raw RMM
 ensemble forecasts of 1-day accumulated precipitation over West
 Sahel in monsoon season 2013, verified against station observations.
 Same setup as in Fig.~\ref{fig:ECMWF24.PIT}.
\label{fig:TIGGE.PIT}}
\end{figure}

\begin{figure}[pt]      
\centering \includegraphics[width = \textwidth]{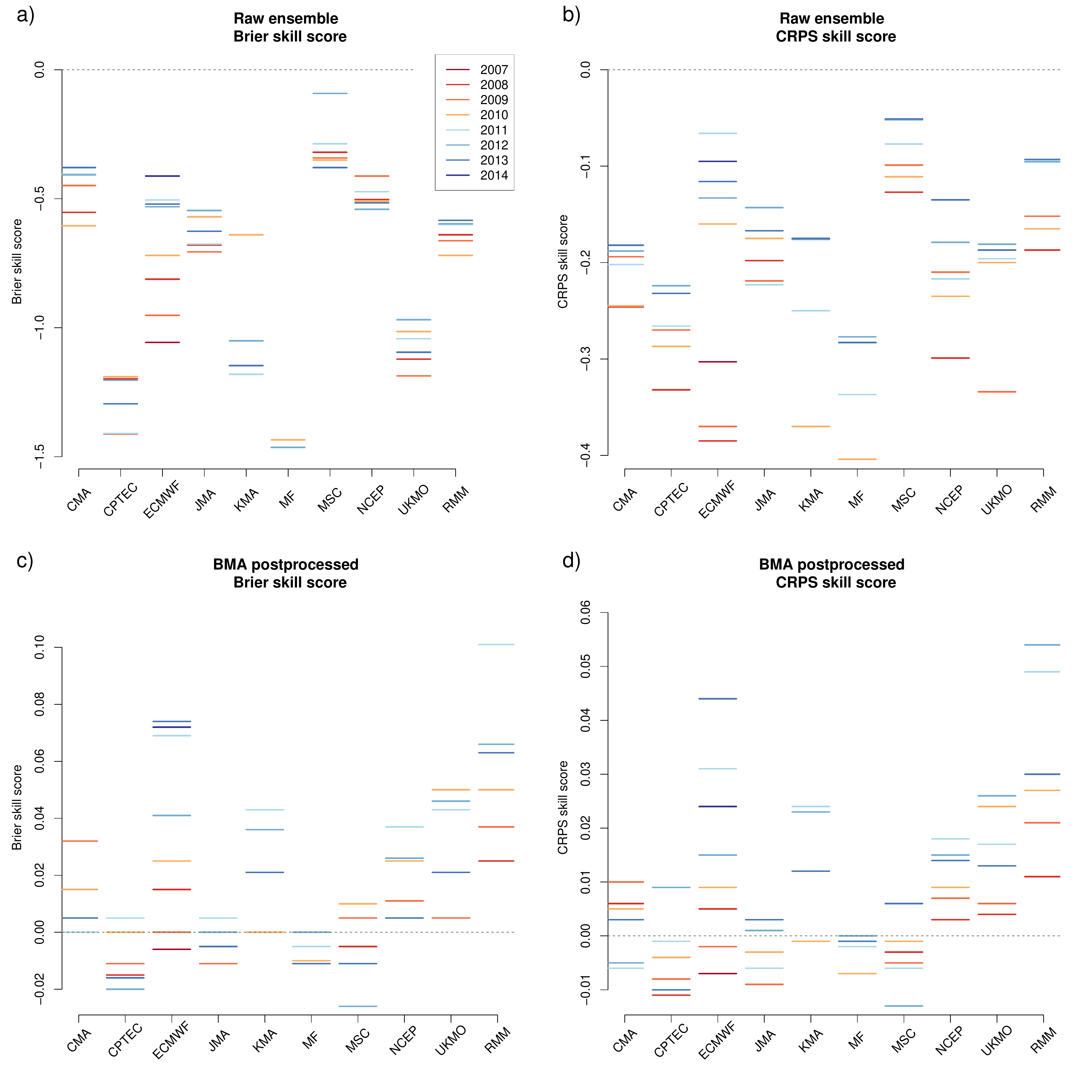}
\caption{Brier and CRPS skill scores for raw and BMA postprocessed
 TIGGE sub-ensemble forecasts of 1-day accumulated precipitation over
 West Sahel in monsoon seasons 2007--2014, verified against station
 observations.  Skill equal to EPC is indicated by the dashed
 line.  \label{fig:TIGGE.scores}}
\end{figure}

\subsection{TIGGE sub-ensembles and RMM ensemble}  \label{subsec:TIGGE}

In addition to the ECMWF EPS, which we have studied thus far, the
TIGGE database contains several more operational sub-ensembles, as
listed in Table \ref{tab:TIGGE}.  Figure \ref{fig:TIGGE.PIT} shows PIT
histograms for the various sub-ensembles and the reduced multi-model
(RMM) ensemble for 1-day accumulated precipitation forecasts over West
Sahel.  All TIGGE sub-ensembles exhibit underdispersion and wet
biases, though in strongly varying degrees.

\begin{figure}[h]
\centering
\includegraphics[width = 0.45 \textwidth]{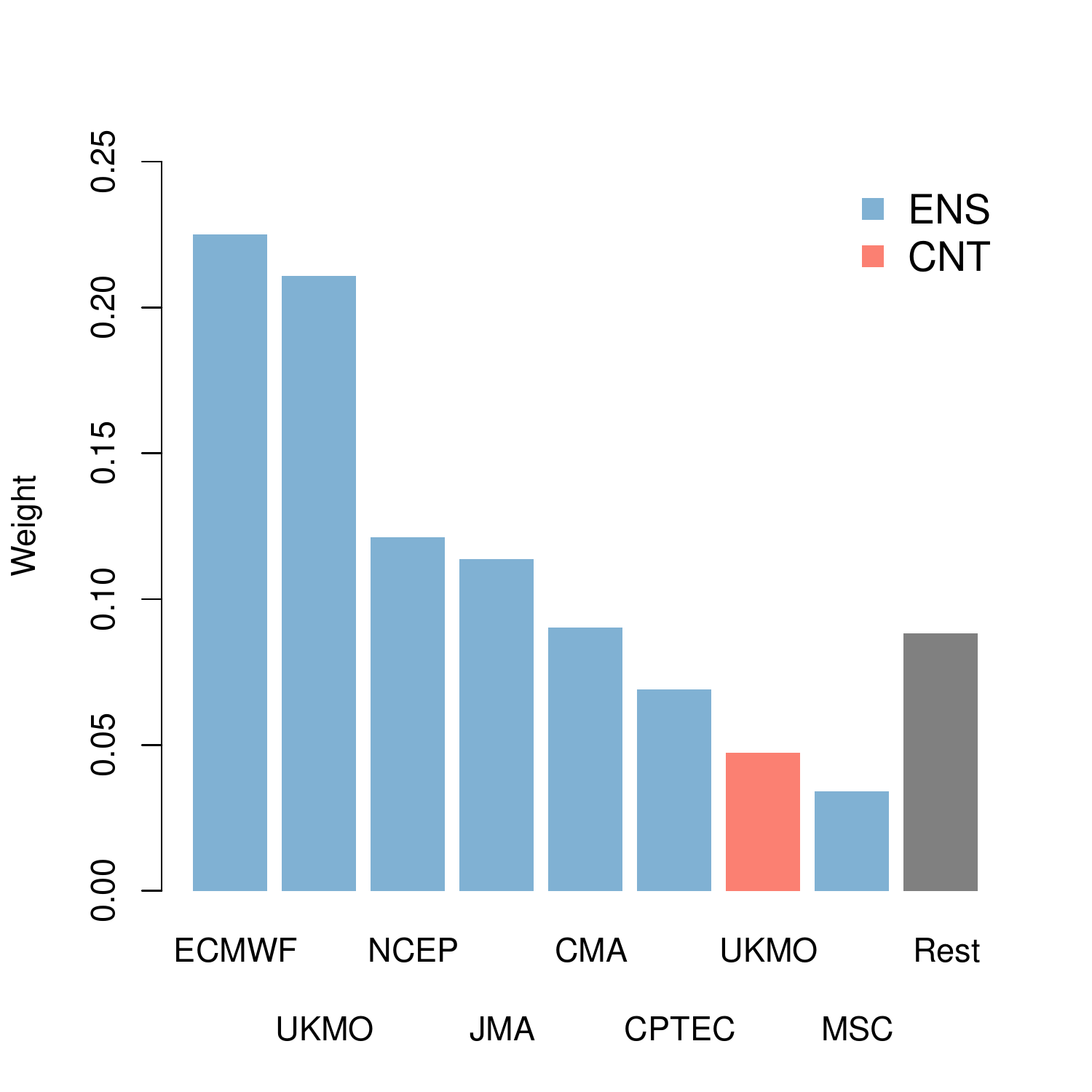}
\caption{BMA weights of RMM components for 1-day accumulated
 precipitation forecasts over West Sahel trained against station
 observations, averaged over monsoon seasons 2008--2013.  Mean
 perturbed forecasts (ENS) and control runs (CNT) are distinguished
 by the color of the respective bar.  \label{fig:RMM}}
\end{figure}

Figure \ref{fig:TIGGE.scores} displays Brier and CRPS skill scores
relative to EPC for raw and BMA postprocessed TIGGE sub-ensemble and
RMM ensemble forecasts in 2007--2014, verified against station
observations.  All raw ensembles underperform relative to EPC, in part
drastically so.  For most sub-ensembles, a temporal improvement in
skill is visible, with monsoon seasons 2011--2014 revealing higher
skill than 2007--2010.  Postprocessing by BMA increases forecast
quality.  The ECMWF, KMA, NCEP, and UKMO ensembles yield the best
postprocessed forecasts, exhibiting small positive skill relative to
EPC for most monsoon periods.  The BMA postprocessed RMM ensemble
outperforms all sub-ensembles as well as EPC, but the improvement is
small.  As shown in Fig.~\ref{fig:RMM}, the mean perturbed forecasts
from the ECMWF, UKMO, and NCEP ensembles are the top three
contributors to the BMA postprocessed RMM forecast.

In further experiments, we have studied raw and postprocessed TIGGE
sub-ensemble and RMM ensemble forecasts at accumulation times up to 5
days and spatial aggregations up to $5^\circ \times 2^\circ$ gridboxes
in TRMM.  Our findings generally remain unchanged.  The raw ensemble
forecasts never reaches the quality of the climatological reference
EPC.  After postprocessing with BMA, the ECMWF ensemble typically
becomes the best performing TIGGE sub-ensemble, showing slightly
better scores than EPC when verified against TRMM observations, at all
spatial aggregations.  The BMA postprocessed RMM forecast depends
heavily on the ECMWF mean perturbed forecast, and is superior to both
EPC and the BMA postprocessed sub-ensemble.

\section{Discussion}  \label{sec:Discussion}

In a first-ever thorough verification study, the quality of
operational ensemble precipitation forecasts from different NWP
centers was assessed over northern tropical Africa for several years,
accumulation periods, and for station and spatially aggregated
satellite observations.  All raw ensembles exhibit calibration
problems in form of underdispersion and biases, and are unreliable at
high PoP forecast values.  They have lower skill than the
climatological reference EPC for the prediction of occurrence and
amount of precipitation, with the underperformance being stable across
monsoon seasons.

After correcting for systematic errors in the raw ensemble through
statistical postprocessing, the ensemble forecasts become reliable and
calibrated, but only few are slightly superior to EPC.  Not
surprisingly, forecast skill tends to be highest for long accumulation
times and large spatial aggregations.  Overall, raw ensemble forecasts
are of no use for the prediction of precipitation over northern
tropical Africa, and even EMOS and BMA postprocessed forecasts have
little added value compared to EPC.  

What are the reasons for this rather disappointing performance of
state-of-the-art global EPSs?  For 1-day accumulated precipitation
forecasts, the ability of an NWP model to resolve the details of
convective organization is essential.  As all global EPSs use
parameterized convection, this clearly limits forecast skill.  The
fact that even postprocessed 1-day accumulated ensemble forecasts
exhibit no skill relative to EPC, implies that ensembles cannot
translate information on the current atmospheric state (e.g., tropical
waves or influences from the extratropics) into meaningful impacts
regarding the occurrence or amount of precipitation.  This is robust
for verification against station as well as satellite observations,
and can therefore not be explained by propagation errors.

For longer accumulation times and larger spatial aggregations, the
large-scale circulation has a much stronger impact on convective
activity, which should weaken the limitation through convective
parameterization.  The skill of 5-day accumulated precipitation
forecasts, however, increases only slightly, if at all, compared to
1-day accumulated forecasts.  The most likely reason for this is that
squall lines have feedbacks on the large-scale circulation, which are
not realistically represented in global NWP models either.
\cite{Marsham2013} find that the large-scale monsoon state in (more
realistic) simulations with explicit convection differs quite
pronouncedly from runs with parameterized convection, even when using
the same resolution of 12 km.  In the explicit-convection simulation,
greater latent and radiative heating to the north weakens the monsoon
flow, delays the diurnal cycle, and convective cold pools provide an
essential component to the monsoon flux.  We suspect that some or all
of these effects are misrepresented in global EPS forecasts.  

The fact that EPS precipitation forecasts are so poor over northern
tropical Africa is a strong demonstration of the complexity of the
underlying forecast problem.  An interesting question in this context
is whether poor predictability in the Tropics is unique to northern
Africa with its strongly organized, weakly synoptically forced
rainfall systems.

Furthermore, the lack of skill motivates alternative approaches to
predicting precipitation over this region.  \cite{Little2009} compare
operational NCEP ensemble, climatological, and statistical forecasts
for stations in the Thames Valley, United Kingdom.  They note that
NCEP forecasts outperform climatological forecasts, but demonstrate
that statistical forecasts, solely based on past observations, can
outperform NCEP forecasts by exploiting spatio-temporal dependencies.
These also exist over northern tropical Africa and some additional
predictability may stem from large-scale drivers such as
convectively-coupled waves.  \cite{Fink2003} note a coupling of the
initiation of squall lines to African easterly waves and
\cite{Wheeler1999} the influence of large-scale tropical waves, such
as Kelvin and equatorial Rossby waves or the Madden-Julian
oscillation, on convective activity. \cite{Pohl2009}
 confirm the relation between the Madden-Julian oscillation and
 rainfall over West Africa and \cite{Vizy2014} demonstrate an impact
 of potential extratropical wave trains on Sahelian
 rainfall. Statistical models based on spatio-temporal
characteristics of rainfall and extended by such large-scale
predictors seem a promising approach to improve precipitation
forecasts over our study region, and we expect such forecasts to
outperform climatology.  This approach will be explored in future
work.

Alternatively, ensembles of convection-permitting NWP model runs,
ideally in combination with ensemble data assimilation, could be used,
but the computational costs are high, and it will take time until a
multi-year database will become available for validation studies.
Given the growing socio-economic impact of rainfall in northern
tropical Africa with its rain-fed agriculture, statistical and
statistical-dynamical approaches should be fostered in parallel in
order to improve the predictability of rainfall in this region.

\section*{Acknowledgments}
The research leading to these results has been accomplished within
project C2 ``Prediction of wet and dry periods of the West African
Monsoon'' of the Transregional Collaborative Research Center SFB / TRR
165 ``Waves to Weather'' funded by the German Science Foundation
(DFG).  Tilmann Gneiting is grateful for support by the Klaus Tschira
Foundation.  The authors also thank various colleagues and weather
services that have over the years contributed to the enrichment of the KASS-D
database; special thanks go to Robert Redl for creating the underlying
software.  Alexander Jordan and Michael Scheuerer have kindly provided
{\tt R} code, and we are grateful to Sebastian Lerch for constructive 
suggestions.

\section*{Appendix}
Rainfall exhibits extremely high spatial and temporal variability,
which hinders automated quality checks applicable to other
meteorological variables such as temperature or pressure.  For
precipitation, \cite{Fiebrich2001} note only a range and a step test.
The global range of station observed 1-day accumulated precipitation
is from 0 mm to 1,825 mm. All KASS-D observations passed this test.
The step test checks if the difference of neighboring 5-minute
accumulated precipitation is smaller than 25 mm.  For 1-day
accumulated precipitation tests of this type are not meaningful, nor
are the persistence tests used by \cite{Pinson2012} for wind speed.

However, the site-specific climatological distributions of
precipitation accumulation should be right-skewed, i.e., the median
should be smaller than the mean, and in the Tropics they should have a
point mass at zero \citep{Rodwell2010}.  As noted, we only consider
stations with more than 80\% available observations in any of the
monsoon seasons, and all 132 stations thus selected passed these
tests.

\bibliographystyle{ametsoc2014}
\bibliography{references}

\end{document}